\documentclass[twocolumn]{jetpl}

\usepackage{amsmath}
\usepackage{amssymb}
\usepackage{amsfonts}
\usepackage{mathrsfs}
\usepackage{graphicx}
\usepackage{color}
\usepackage{bm}
\usepackage{cite}

\begin{document}

\title{Mesoscopic fluctuations of the local density of states in interacting electron systems}

\rtitle{Mesoscopic fluctuations of the local density of states in interacting electron systems}

\sodtitle{Mesoscopic fluctuations of the local density of states in interacting electron systems}

\author{I.\,S.\,Burmistrov$^{1}$\thanks{e-mail: burmi@itp.ac.ru.},  I.\,V.\,Gornyi$^{2,3,4}$, A.\,D.\,Mirlin$^{2,4,5}$}

\dates{\today}{*}

\address{
$^{1}$ L. D. Landau Institute for Theoretical Physics RAS, 119334 Moscow, Russia \\
$^{2}$ Institut f\"ur Nanotechnologie, Karlsruhe Institute of Technology, 76021 Karlsruhe, Germany \\
$^{3}$ A. F. Ioffe Physico-Technical Institute, 194021 St. Petersburg, Russia \\
$^{4}$ Institut f\"ur Theorie der kondensierten Materie, Karlsruhe Institute of Technology, 76128 Karlsruhe, Germany \\
$^{5}$ Petersburg Nuclear Physics Institute, 188300 St. Petersburg, Russia
}

\abstract{We review our recent theoretical results for mesoscopic fluctuations of the local density of states in the presence of electron-electron interaction. We focus on the two specific cases: (i) a vicinity of interacting critical point corresponding to Anderson-Mott transition, and (ii) a vicinity of non-interacting critical point in the presence of a weak electron-electron attraction. In both cases  strong mesoscopic fluctuations of the local density of states exist.
}

\maketitle

\textsf{Introduction.}\ --- Since the seminal paper by P.W. Anderson \cite{Anderson58} a study of the localization-delocalization quantum phase transition in noninteracting disordered systems has turned into a vast field of research (see Refs. \cite{AL50,Evers08} for a review). As any other quantum phase transitions, Anderson transition is characterized by a set of critical exponents which controls the scaling of a divergent correlation length and different physical observables. However, contrary to ordinary quantum phase transitions, for an Anderson transition there is the additional set of critical exponents $\Delta_q^{\rm n}$ which determines the scaling behaviour of electron wave functions. Specifically, the disorder-averaged $q$-th moment of an electron wave function (the inverse participation ratio) has the multifractal behaviour at criticality \cite{Wegner1980,Castellani1986}:
\begin{equation}
\langle P_q\rangle=\int\limits_{r<L} d^d\bm{r} \left \langle\,\,  \bigl |\psi(\bm{r})\bigr |^{2q}\right \rangle 
\sim L^{-d(q-1) - \Delta_q^{\rm n}} .
\label{eq:Psi}
\end{equation}
Here $L$ stands for a system size and $\langle \cdots \rangle$ denotes the averaging over disorder. The multifractal exponents $\Delta_q^{\rm n} \leqslant 0$ are nonlinear functions of $q$. [Here and below the superscript (as well as subscript) `${\rm n}$' is used for quantities related to the non-interacting critical point.] Eq. \eqref{eq:Psi} implies the existence of strong mesoscopic fluctuations of wave functions. 

Multifractality in non-interacting disordered systems has been remaining for a long time a concept which was studied either theoretically or in numerical simulations (see Refs. \cite{Janssen,Huckestein,Evers08} for a review). Recently, multifractality has become a subject of experimental research. For example,  an indication of multifractality has been reported in scanning tunneling spectroscopy data in diluted magnetic semiconductor Ga$_{1-x}$Mn$_x$As \cite{Richardella}, in experimental studies of ultrasound waves propagating through the system of randomly packed Al beads \cite{Faez2009}, and in experimental data on spreading of light waves in the dielectric nanoneedle array \cite{Mascheck2012}. 

Multifractality of wave functions \eqref{eq:Psi} can be formulated equivalently as the following scaling behaviour of the moments of the local density of states \cite{Lerner1988}:
\begin{equation}
\langle \rho^q(E,\bm{r})\rangle \propto L^{-\Delta_q^{\rm n}} .
\label{eq:RR1}
\end{equation}
Relations  \eqref{eq:RR1} are remarkable not only due to the fact that $\Delta_q^{\rm n}\leqslant 0$ but also for the following reason. Typically, one expects existence of subleading corrections to Eq. \eqref{eq:RR1}. Such corrections to scaling are completely absent for the moments of  the local density of states. In fact, it is known \cite{Wegner1986,Wegner1987a,Wegner1987b} that many more correlation functions of electron wave functions should demonstrate pure scaling behaviour with non-positive critical exponents similar to Eq. \eqref{eq:RR1}. Recently, one of us, with Gruzberg and Zirnbauer, proposed a method to construct all such pure scaling observables in terms of disorder-averaged combinations of electron wave functions and demonstrated that their critical exponents obey a set of exact symmetry relations \cite{Gruzberg}. Mesoscopic fluctuations of electron wave functions have interesting consequences for the Kondo problem \cite{Kettemann2006,Micklitz2006,Kettemann2007} and the Anderson orthogonality catastrophe \cite{Kettemann2016}.

The progress in theoretical understanding of multifractality at Anderson transitions has been achieved for disordered systems without electron-electron interactions. As is well-known, a metal-insulator transition can survive in the presence of electron-electron interaction \cite{McMillan,Finkelstein1983a,Castellani1984}. In this case the metal-insulator transition is usually termed as Anderson-Mott (or Mott-Anderson) transition (see Refs. \cite{Finkelstein1990,BelitzKirkpatrick1994} for a review). In most cases, a non-interacting critical point is unstable  towards electron-electron interaction in the renormalization group sense. The authors are aware of the only exception when non-interacting Anderson transition survives in the presence of electron-electron interaction.  This is the case of broken time reversal and spin-rotational symmetries and a short-range electron-electron interaction. In this situation the non-interacting multifractal exponents determine scaling of the interaction-induced dephasing at criticality \cite{HW,WFGC,BBEGM}.

\emph{Two scenarios} are possible if a non-interacting critical point is unstable with respect to interaction. In the first scenario there exists
an unstable critical point at finite value of electron-electron interaction which separates metallic and insulating phases. This Mott-Anderson transition is characterized by critical exponents which are different from critical exponents in a non-interacting case. In the second scenario the electron-electron interaction results in the instability, e.g. superconducting instability or Stoner instability, at finite renormalization group scale. Then a much more complicated phase diagram arises than in the absence of interaction.

In the first scenario, i.e. at Anderson-Mott transition, formulation of multifractality in terms of moments of electron wave functions, Eq. \eqref{eq:Psi}, loses its significance. However, scaling of moments of the local density states at interacting criticality is well posed question. A fate of mesoscopic fluctuations of $\rho(E,\bm{r})$ in the first scenario, i.e. at an Anderson-Mott transition, has been not explored until recently. Attempts to address this question have been performed in 
numerical analysis of disordered electrons with Coulomb interaction in the framework of functional density theory \cite{Slevin2012,Slevin2014} and by numerical implementation of the Hartree-Fock scheme \cite{Amini2014}. 
Recently, the detailed theory of mesoscopic fluctuations of the local density of states has been developed by the present authors within nonlinear sigma model treatment of disordered interacting electrons in $d=2+\epsilon$ dimensions \cite{Burmistrov2013,Burmistrov2014,Burmistrov2015a}. It was demonstrated that moments of the local density of states 
at Mott-Anderson transitions behave generically similar to Eq. \eqref{eq:RR1} although the corresponding critical exponents are different from their non-interacting counterparts $\Delta_q^{\rm n}$. 

The second scenario with instability due to attractive interaction in the Cooper channel has been in the focus of theoretical research during last decade\cite{FeigelmanYuzbashyan2007,FeigelmanCuevas2010,BurmistrovGornyiMirlin2012,Foster2012,DellAnna2013,Foster2014,Burmistrov2015b,DellAnna2017}. It was found that in some range of parameters multifractality favours the superconducting instability which results in enhanced superconducting transition temperature $T_c$ in comparison with the clean case. Recently, the present authors have demonstrated that near the superconducting transition with enhanced $T_c$ one can expect enhanced mesoscopic fluctuations of the local density of states governed by the critical exponents for the non-interacting critical point \cite{Burmistrov2016}. 

In this brief review we discuss the mesoscopic fluctuations of the local density of states for the two scenarios. The review is based on the results published recently by the present authors \cite{Burmistrov2013,Burmistrov2014,Burmistrov2015a,Burmistrov2016}.

\textsf{Scaling near interacting critical point.}\ --- We start from discussion of a general scaling behaviour of moments of the local density of states near an interacting critical point, $t=t_*$. Here $t$ stands for the dimensionless resistance which is related with the dimensionless conductance $g$ measured in units $e^2/h$: $t=2/\pi g$. This critical point describes the Mott-Anderson transition between metallic and insulating phases and is characterized by the divergent correlation/localization length $\xi= l |1-t/t_*|^{-\nu}$. Here $l$ denotes the mean free path.

It is worthwhile to mention that a fate of multifractality in the local density of states in the presence of electron-electron interaction is by no means obvious. The reason is the phenomenon of strong suppression of the disorder-averaged local density of states at the Fermi energy. At strong disorder this suppression is known as Coulomb gap \cite{ES75,SE84} whereas at weak disorder it is the so-called zero-bias anomaly \cite{AA1979,AAL1980,AAL1980b,Fin198384a,Fin198384b,nazarov89,levitov97,kamenev99}. 
The evolution of the zero-bias anomaly into Coulomb gap across the Anderson-Mott transition was intensively studied experimentally \cite{Mochel,Massey,lee,Teizer,lee2004,aubin}. 

At a first glance, the suppression of the disorder-averaged local density of states should prevent from multifractal behaviour of its moments. However, similar situation is known to occur in non-interacting systems of fermions in symmetry class C where in spite of power-law suppression of the average local density of states at zero energy, its moments behave multifractally \cite{Evers08}. 

The zero bias anomaly translates into the following scaling behaviour of the average local density of states at the Fermi energy and temperature, $E=T=0$, at criticality (we note that we count the energy from the Fermi energy)\cite{Finkelstein1990,BelitzKirkpatrick1994},
\begin{equation}
\langle \rho \rangle \sim (\xi/l)^{-\theta} {\Upsilon}(\xi/L) .
\label{Scal_eq2}
\end{equation}
The scaling function ${\Upsilon}$ has the following asymptotic behaviour:
 \begin{equation}
{\Upsilon}(y) = \begin{cases}
1, & \qquad y\ll 1 , \\
y^{\theta}, & \qquad y\gg 1 .
\end{cases} 
\end{equation}
The critical exponent $\theta$ is determined by the anomalous dimension $\zeta$ of $\langle \rho\rangle$ at the critical point, $\theta=\zeta^*$ (see below). Suppression of the local density of states corresponds to $\theta>0$. 
In Refs. \cite{Burmistrov2013} it was shown that  the $q$-th moment of the local density of states  at $E=T=0$ obeys the following scaling law:
\begin{equation}
\langle \rho^q \rangle = \langle \rho \rangle^q \left ({\xi}/{l}\right )^{-\Delta_q} \Upsilon_q(\xi/L) .
\label{Scal_eq3}
\end{equation}
The scaling function $\Upsilon_q(y)$ has asymptotes at $y\ll 1$ and $y\gg 1$ similar to the function $\Upsilon(y)$:
 \begin{equation}
{\Upsilon_q}(y) = \begin{cases}
1, & \qquad y\ll 1 , \\
y^{\Delta_q}, & \qquad y\gg 1 .
\end{cases} 
\end{equation}
 The multifractal critical exponent $\Delta_q$ is determined by the anomalous dimension $\zeta_q$ of the $q$-th moment of the normalized local density of states, $\langle \rho^q \rangle/\langle \rho \rangle^q$,  at the critical point, $\Delta_q = \zeta_q^*$ (see below). 
 
Definition of $\Delta_q$ via the moments of the local density of states is obviously limited to integer positive values of $q$. However, similar to the noninteracting case, $\Delta_q$ can be extended (by analytic continuation) to all real (and, in fact, even complex) $q$. This is possible since the local density of states is a real positive quantity. We note that the multifractal exponents obey the same general properties as in the non-interacting case:
$\Delta_0=\Delta_1=0$ and $d^2\Delta_q/dq^2<0$. 
This implies that $\Delta_q>0$ ($\Delta_q\leqslant 0$) for $0<q<1$ (otherwise).

At finite energy or temperature electron-electron interaction induces the inelastic length $L_\phi$ related with the  dephasing time $\tau_\phi$ via a dynamical exponent $z$, $L_\phi \sim \tau_\phi^{1/z}$. We remind that for  Coulomb interaction the scaling with frequency/energy and temperature is the same such that $1/\tau_\phi \sim \max\{|E|,T\}$ and $L_\phi \sim \min\{L_E, L_T\}$, where $L_E\sim |E|^{-1/z}$ and $L_T\sim T^{-1/z}$. For $L_\phi \ll L$ the inelastic length plays the role of the effective system size. Therefore, at finite energy and temperature Eq.~\eqref{Scal_eq3} becomes as follows
\begin{equation}
\langle \rho^q(E,\bm{r}) \rangle  \sim \langle \rho(E) \rangle^q \bigl ({\mathcal{L}}/{l}\bigr )^{-\Delta_q} \sim \bigl ({\mathcal{L}}/{l}\bigr )^{-\theta q-\Delta_q} ,
\label{Scal_eq5}
\end{equation}
where $\mathcal{L} = \min\{L,\xi,L_\phi\}$. The sign of the exponent $\theta q+\Delta_q$ depends on the value of $q$. Since $\theta>0$, $\theta q+\Delta_q$ remains positive for not too large positive $q$. The absolute value of $\Delta_q$ is expected to grow sufficiently fast (typically, as $q^2$). Thus $\theta q+\Delta_q$ becomes negative for large enough $q$. Therefore, we have a counterintuitive behaviour of the local density of states as opposed to a clean system: its average value is suppressed whereas its sufficiently high moments are enhanced. This occurs due to  a combined effect of interaction and disorder. 

The nontrivial scaling of moments of the local density of states translates into scaling behaviour of frequency and spatial dependence of its correlation functions. For example, the 2-point correlation function of the local density of states becomes
\begin{gather}
\langle \rho(E,\bm{r}) \rho(E+\omega,\bm{r}+\bm{R})\rangle 
\sim\langle \rho(E) \rangle^2 \hspace{2cm}{\,}\notag \\
\times \begin{cases}
(\mathcal{L}/L_\omega)^{\theta} (L_\omega/R)^{-\Delta_2}, & \quad R\ll L_\omega \ll \mathcal{L} ,\\
(\mathcal{L}/L_\omega)^{\theta}, & \quad L_\omega \ll R, \mathcal{L} , \\
(\mathcal{L}/R)^{-\Delta_2}, & \quad R \ll  \mathcal{L} \ll  L_\omega, \\
1, & \quad \mathcal{L} \ll  R, L_\omega .
\end{cases}
\label{eq:Scal:2pLDOS}
\end{gather}

It is instructive to discuss the behaviour of this 2-point correlation function at zero temperature and frequency, $\omega=T=0$, and  in the infinite system size limit, $L\to \infty$. Exactly at the critical point, $t=t_*$, the correlation length diverges, $\xi=\infty$, and the scale $\mathcal{L} \equiv L_E$. Then, Eq. \eqref{eq:Scal:2pLDOS} implies that the multifractal correlations persist upto $R\sim L_E$, i.e. they become long-ranged near the Fermi energy, $E=0$. At criticality on the metallic side of the transition, $t<t_*$, the multifractal correlations exist up to the spatial scale $\min\{\xi,L_E\}$. The competition between $\xi$ and $L_E$ determines the energy scale $\Delta_\xi \sim \xi^{-1/z} \sim (t_*-t)^{\nu z}$ which controls the critical region near the interacting critical point. Near the Fermi level, $|E|<\Delta_\xi$, the spatial extent of the multifractal correlations is controled by $\xi$,  i.e. the correlations become effectively short-ranged. Away from the Fermi energy, $|E|>\Delta_\xi$, the multifractal correlations exist upto $R \sim L_E < \xi$. 

\begin{figure}[t]
\centerline{\includegraphics[angle=0,width=0.9\columnwidth]{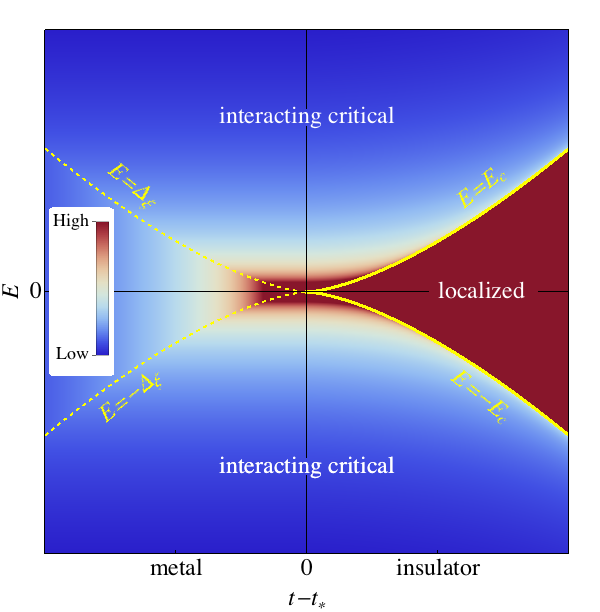}}
\caption{Fig. \protect\ref{Figure2}. Sketch of the phase diagram in the energy ($E$) vs disorder ($t$) plane. 
The interacting critical point is situated at $t=t_*$ and zero energy, $E=0$. The dashed yellow curves separate critical regime on metallic side. The solid yellow curves correspond to the mobility edge $E=\pm E_c$. The colour scheme indicates value of  
the ratio $\langle \rho^2(E)\rangle /\langle \rho(E)\rangle^2$ in different parts of the phase diagram.
}
\label{Figure2}
\end{figure}

On the insulating side of the criticallity the situation can be more complicated. We assume that Anderson transition occurs in the absence of interaction at $t=t_*^{\rm (n)} > t_*$, i.e. interaction favours localization.  The complication arises from the dependence of localization length $\xi(E)$ on an excitation energy $E$. The normalization is such that $\xi(E=0)=\xi$. We remind that the dependence of localization length on energy is natural for the Anderson transition in the absence of interaction due to the existence of the mobility edge, $E_c^{\rm (n)} \sim t-t_*^{\rm (n)}$. In the case of the interacting critical point, we demonstrated \cite{Burmistrov2014} that the mobility edge $E_c$ exists for single-particle excitations (particles and holes) but it scaling with the distance to the critical point differs from the non-interacting case:
\begin{equation}
E_c =  \Delta_\xi \left (\frac{t_*^{\rm (n)}-t_*}{t_*^{\rm (n)}} \right )^{\nu z} \propto  \left (\frac{t-t_*}{t_*} \right )^{\nu z} \, .
\label{eq:IntMobEdge}
\end{equation}
The single particle excitations are localized (delocalized) for $|E|<E_c$  ($|E| > E_c$). The overall  phase diagram of the Anderson-Mott transition discussed above is sketched in Fig.~\ref{Figure2}. 
The localization length $\xi(E)$ scales near $E_c$ as
\begin{equation}
 \label{eq:NILocLength}
 \xi(E) = \xi \left (\frac{E_c-|E|}{E_c}\right )^{-\nu_{\rm n}}\,, \quad E_c-|E|  \ll E_c \,,
\end{equation}
where $\nu_{\rm n}$ is the exponent for the corresponding non-interacting Anderson transition. The zero-temperature dephasing rate vanishes at $|E|<E_c$ and demonstrates critical behaviour for $|E| > E_c$:
\begin{equation}
\label{eq:NIDepLength}
L_\phi = L_E \begin{cases}
 \left (\frac{|E|-E_c}{E_c}\right )^{-1/z_\phi^{\rm n}}, & \quad
|E|-E_c \ll E_c , \\
1, & \quad |E| \gg E_c .
\end{cases}
\end{equation}
The critical behaviour of  $L_\phi$ near the mobility edge occurs since the zero temperature decay is only possible in the continuous spectrum. However, the corresponding phase volume tends to zero as $E$ approaches $E_c$ from
above. An estimate based on the Fermi golden rule yields $z_\phi^{\rm n} = \max\{d^2/(4-d),d^2/(d+\Delta_2^{\rm n})\}$ \cite{Burmistrov2014}. 

Derivation of Eq. \eqref{eq:IntMobEdge} is valid under assumption $ \nu z > 1 $, which allows us to neglect the energy dependence of bare diffusion coefficient (included in $t$) on energy. We note that the condition  $ \nu z > 1 $ holds for many examples of Anderson-Mott transitions in the presence of interaction. 

We note that a phase diagram qualitatively similar to the phase diagram shown in Fig. \ref{Figure2} was obtained in Ref.  \cite{Amini2014}
based on the numerical modeling within Hartree-Fock wave functions of the Mott-Anderson transition on a 3D cubic
lattice of linear size $10$.

On the insulating side of criticality the mesoscopic fluctuations of the local density of states are governed by competition between  multifractality at the interacting and non-interacting critical points as well as by localization
of single-particle excitations for $|E|<E_c$. At $|E|\gg E_c$ the presence of the mobility edge is immaterial
and the moments of the local density of states behave in the same way as on the metallic side of the criticality:
\begin{equation}
 \langle \rho^q(E) \rangle \sim  \langle \rho(E) \rangle^q L_{\phi}^{-\Delta_q}\,.
\end{equation}
In the vicinity of the mobility edge from above,  $0< |E|-E_c\ll
E_c$, the mesoscopic fluctuations of $\rho(E,\bm{r})$ are further enhanced due to non-interacting critical behaviour near $E_c$:
\begin{equation}
 \langle \rho^q(E) \rangle \sim  \langle \rho(E) \rangle^q \xi^{-\Delta_q} (L_\phi/\xi)^{-\Delta_q^{\rm n}}\,.
\end{equation}
Since for $|E|<E_c$, the zero-temperature dephasing length is infinite and the mesoscopic fluctuations of the local density of states are controlled by the system size $L$. Nonzero temperature induces a finite (albeit large) dephasing length $L_{\phi T}$. 
Under the assumption that $L \gg L_{\phi T} \gg \xi(E)$ we find for $|E| < E_c$:
\begin{gather}
 \langle \rho^q(E) \rangle \sim  \langle \rho(E) \rangle^q 
 \xi^{-\Delta_q} \left (\frac{L_{\phi T}}{\xi}\right )^{d(q-1)}\hspace{2cm}{
 \,} \notag \\
\times  \begin{cases}
 \left ({\xi(E)}/{\xi}\right )^{-\Delta_q^{\rm n}-d(q-1)} , & \, 0< E_c-|E| \ll E_c ,\\
  1 , & \, 0< E_c-|E| \sim E_c .
\end{cases}
\end{gather}

The presence of the mobility edge at $E=\pm E_c$ affects also the spatial correlations of the local density of states on the insulating side of the interacting criticality. Since at $|E|\gg E_c$ the system is controlled by interacting critical point 
the 2-point correlation function of the local density of states obeys the following power-law scaling for $R \ll L_\phi$:
\begin{equation}
\langle\rho(E,\bm{r})\rho(E,\bm{r}+\bm{R})\rangle
\sim \langle\rho(E)\rangle^2 \bigl (R/L_\phi\bigr )^{\Delta_2} .
\end{equation}
In the vicinity of the mobility edge from above, $0<|E|-E_c\ll E_c$, there are the interacting multifractal scaling of the 2-point correlation function upto the scale $\xi$ and the non-interacting multifractal scaling for $\xi\ll R\ll L_\phi$:
\begin{gather}
{\,}\hspace{-3cm} \langle\rho(E,\bm{r})\rho(E,\bm{r}+\bm{R})\rangle
\sim \langle\rho(E)\rangle^2 \notag \\
\times 
\begin{cases}
\bigl (R/\xi\bigr )^{\Delta_2} \bigl (\xi/L_\phi\bigr )^{\Delta_2^{\rm n}} , & \, R\ll \xi ,\\
\bigl (R/L_\phi\bigr )^{\Delta_2^{\rm n}} ,  & \, \xi \ll R \ll L_\phi  .
\end{cases}
\end{gather}
For energies below but close to $E_c$, $0<E_c-|E|\ll E_c$,  the system shows first the interacting multifractal scaling up to the scale $\xi$, then the non-interacting multifractality up to the scale $\xi(E)$, and finally, insulator-like fluctuations up to the scale $L_{\phi T}$:
\begin{gather}
{\,}\hspace{-3cm} \langle\rho(E,\bm{r})\rho(E,\bm{r}+\bm{R})\rangle
\sim \langle\rho(E)\rangle^2 \notag \\
\times 
\begin{cases}
\bigl (R/\xi\bigr )^{\Delta_2} \bigl (\xi/\xi(E)\bigr )^{\Delta_2^{\rm n}}  \bigl (L_{\phi T}/\xi(E)\bigr )^{d} , \qquad  R\ll \xi ,\\
\bigl (R/\xi(E)\bigr )^{\Delta_2^{\rm n}}  \bigl (L_{\phi T}/\xi(E)\bigr )^{d} , \qquad  \xi \ll R \ll \xi(E) .
\end{cases}
\end{gather}
At energies well below $E_c$, $0<E_c-|E| \sim E_c$, the localization length $\xi(E)$ is of the order of $\xi$. Therefore, for $R\ll \xi$ we find
\begin{gather}
 \langle\rho(E,\bm{r})\rho(E,\bm{r}+\bm{R})\rangle
\sim \langle\rho(E)\rangle^2 
\bigl (R/\xi\bigr )^{\Delta_2}  \bigl (L_{\phi T}/\xi\bigr )^{d}  .
\end{gather}

\begin{figure}[t]
\centerline{(a) \includegraphics[width=0.33\textwidth]{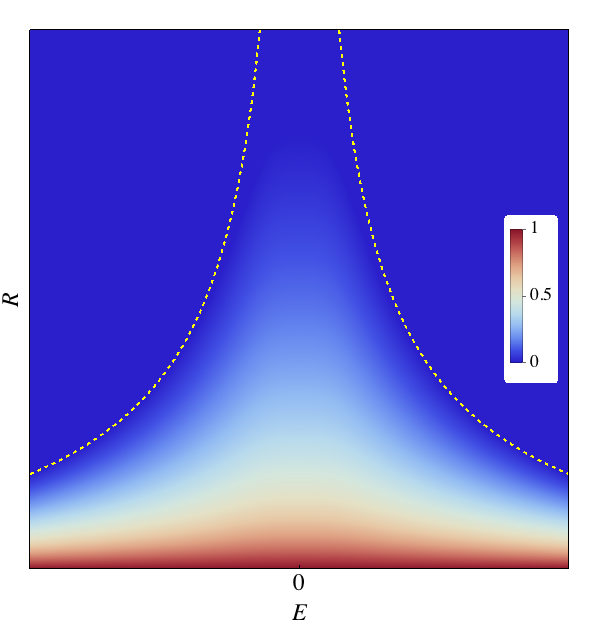}}
\centerline{(b) \includegraphics[width=0.33\textwidth]{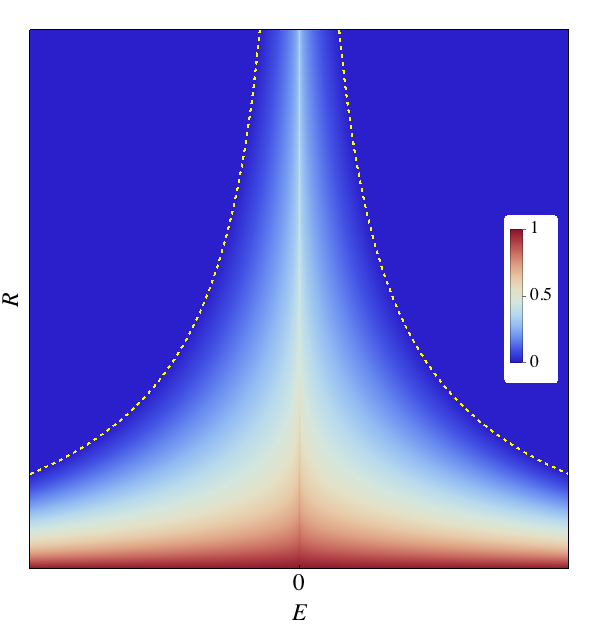}}
\centerline{(c) \includegraphics[width=0.33\textwidth]{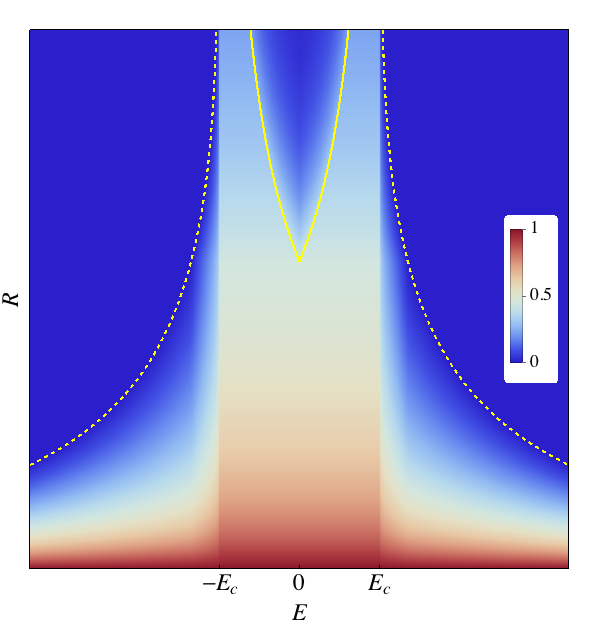}}
\caption{Fig. \protect\ref{Figure1}. Color-code plots illustrating the behaviour of the normalized autocorrelation
function $[\langle\rho(E,\bm{r})\rho(E,\bm{r}+\bm{R})\rangle-\langle\rho(E)\rangle^2]/[\langle\rho^2(E,\bm{r})\rangle-\langle\rho(E)\rangle^2]$ (a) slightly on the metallic side, $t_*<t$, (b) 
at the critical point, $t=t_*$, and (c) slightly on the insulating side, $t>t_*$. The dashed yellow curves in panels (a) and (b) indicate $R = L_E$. In panel (c) the dashed (solid) yellow curves indicate $R=L_\phi$ ($R=\xi(E)$).
}
\label{Figure1}
\end{figure}

Multifractal behaviour of the 2-point correlation function of the local density of states as a function of energy and distance across the Anderson-Mott transition is illustrated in Fig. \ref{Figure1}. We note that qualitatively the behaviour of the 2-point correlation function presented in Figs.  \ref{Figure1}a and  \ref{Figure1}b is consistent with the experimental findings of Ref.~\cite{Richardella}. 

\begin{figure}[t]
(a)\centerline{ \includegraphics[width=0.7\columnwidth]{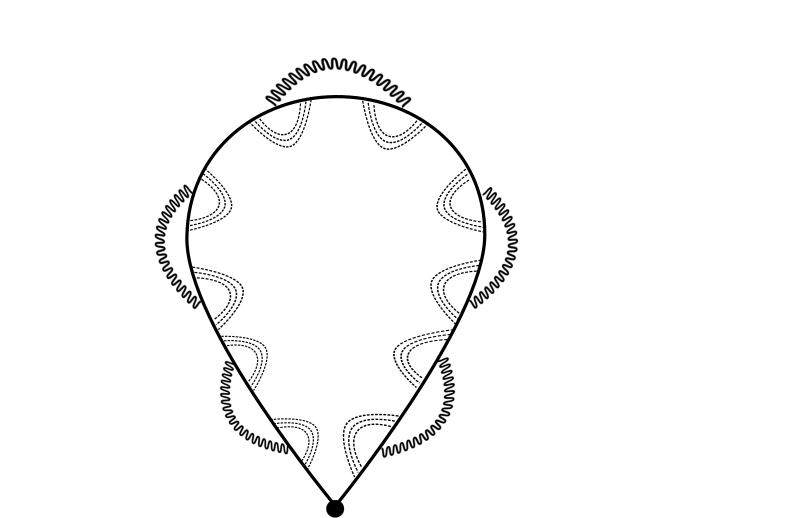}}

(b)\centerline{ \includegraphics[width=0.6\columnwidth]{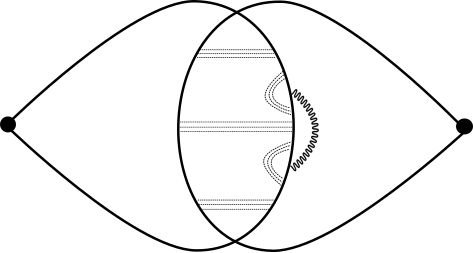}}
\caption{Fig. \protect\ref{Figure3}. Representative diagrams for (a) the disorder-averaged local density of states and (b) two-loop contribution to the second moment of the local density of states. Solid lines denote electron Green functions, while wavy lines denote the dynamically screened Coulomb interaction. Ladders of dashed lines, e.g. dressing the interaction vertices, represent diffusons.
}
\label{Figure3}
\end{figure}

\textsf{Anomalous dimension $\zeta_q$.}\ --- The anomalous dimension that governs the scaling behaviour of moments of the local density of states can be computed near two dimensions as a perturbative expansion in the dimensionless resistance $t$. In the most general case when both time reversal and spin rotational symmetries are preserved we obtained the following result \cite{Burmistrov2015a}:
\begin{equation}
\zeta_q = \frac{q(1-q)}{2} \Bigl [ 2 t + \bigl [c(\gamma_s)+3 c(\gamma_t)-2 \gamma_c \bigr ] \frac{t^2}{2} \Bigr ]+ O(t^3) .
\label{eqmqRG-sym1}
\end{equation}
Here $\gamma_s$, $\gamma_t$, and $\gamma_c$ are dimensionless parameters which describe electron-electron interaction in the singlet and triplet particle-hole channels and in the Cooper channel, respectively. The function $c(\gamma)$ is defined as follows
\begin{gather}
c(\gamma) = 2 +\frac{2+\gamma}{\gamma} {\rm li}_2(-\gamma) + \frac{1+\gamma}{2\gamma} \ln^2(1+\gamma) ,
\label{eq:def:c}
\end{gather}
where ${\rm li}_m(\gamma)=\sum_{k=1}^\infty \gamma^k/k^m$ denotes the polylogarithm.  We remind that the anomalous dimension which controls scaling behaviour of the averaged local density of states is given as follows (see Refs. \cite{Finkelstein1990,BelitzKirkpatrick1994} for a review):
\begin{equation}
\zeta = - \Bigl [ \ln(1+\gamma_s) +3 \ln(1+\gamma_t) +2 \gamma_c\Bigr ] \frac{t}{2} + O(t^2) .
\label{eqmqRG-sym2}
\end{equation}
Typical contributions which lead to the results \eqref{eqmqRG-sym2} and \eqref{eqmqRG-sym1} are shown in Fig. \ref{Figure3}. In the case of broken time reversal and/or spin rotational symmetries the results \eqref{eqmqRG-sym1}  should be modified. The corresponding results are summarized in Table 1. 

%
%
\begin{table*}[t]
\begin{center}
\caption{Table 1. The two-loop results for anomalous dimension $\zeta_q$ derived in Ref. \cite{Burmistrov2015a}. One needs to distinguish 
the cases of spin-orbit coupling (SO) and of the Zeeman splitting (MF) in which the spin-rotational symmetry is broken but in a different way.\label{Tabb1}}
\begin{tabular}{cc|l}
time-reversal sym. & spin-rotational sym. & {anomalous dimension} \\
\hline 
no & no & $\zeta_q = [q(1-q)/2] \bigl [ t/2+c(\gamma_s) t^2/4 \bigr ] +O(t^3)$ \\
no & yes & $\zeta_q = [q(1-q)/2] \bigl [ t+[c(\gamma_s)+3c(\gamma_t)] t^2/2 \bigr ]+O(t^3)$ \\
yes &  no (SO) & $\zeta_q = [q(1-q)/2] \bigl [ t/2+[c(\gamma_s)-2\gamma_c] t^2/4 \bigr ]+O(t^3)$ \\
yes &  no (MF) &$\zeta_q = [q(1-q)/2] \bigl [ t+[c(\gamma_s)+c(\gamma_t)-2\gamma_c] t^2/2 \bigr ]+O(t^3)$ \\
\end{tabular}
\end{center}

\end{table*}

To illustrate our general results we consider the case of the Anderson-Mott transition in $d=2+\epsilon$ dimensions in the presence of Coulomb interaction ($\gamma_s=-1$) and in the absence of time-reversal and spin-rotational symmetries. This situation can be realized in the presence of magnetic impurities. In the case under consideration the dimensionless resistance $t$ is renormalized as follows \cite{baranov02}:
\begin{equation}
-\frac{dt}{dy} = \beta(t) = \epsilon t - t^2 - A t^3 + O(t^4) ,
\label{eq2E1}
\end{equation}
where $y=\ln L/l$ is the running renormalization group scale and 
\begin{gather}
A =\frac{1}{16}\Bigl [\frac{139}{6}+\frac{(\pi^2-18)^2}{12}+\frac{19}{2}\zeta (3)+\Bigl ( 16 + \frac{\pi ^2}{3} \Bigr )\ln ^{2}2  \notag \\
- \Bigl (44-\frac{\pi ^{2}}{2}+7\zeta (3)\Bigr ) \ln 2+16\mathcal{G}-\frac{1}{3}\ln ^{4}2-8{\rm li}_4\left(\frac{1}{2}\right)\Bigr ]\notag \\ \approx 1.64 .
\end{gather}
Here $\zeta(x)$  and $\mathcal{G} \approx 0.915 $ stand for  the Riemann zeta-function and the Catalan constant, respectively.
The critical point $t_*=\epsilon(1-A\epsilon)+O(\epsilon^3)$ follows from the solution of equation $\beta(t_*)=0$. 
In the absence of interaction, the situation we consider corresponds to the unitary Wigner-Dyson class A. In this case the non-interacting $\beta$-function is known up to the five-loop order \cite{Hikami,BW,Wegner89}:
\begin{equation}
-\frac{dt}{d y} = \beta^{({\rm n})}(t) = \epsilon t - \frac{1}{8} t^3 - \frac{3}{128} t^5 + O(t^6) .
\label{eq2E4}
\end{equation}
The non-interacting critical point is given as $t_*^{({\rm n})}=2(2\epsilon)^{1/2}(1-3\epsilon/4)+O(\epsilon^{5/2})$. The critical exponents for the interacting and non-interacting critical points are compared in Table 2. 
 We note that  near two dimensions $t_*^{({\rm n})} \gg t_*$,  therefore, as discussed above,  the mobility edge for the single particle excitations exists at the insulating side of the transition, $t>t_*$.
In the case $d=2+\epsilon$, we find from Eq. \eqref{eq:IntMobEdge} that $E_c = \Delta_\xi \exp(1/\sqrt{2\epsilon}) \gg \Delta_\xi$. Also we note that  the combination $\theta q +\Delta_q$ is positive for $q < 4/\epsilon$.
 For $\epsilon\ll 1$, the expansion in $t$ is parametrically controlled, the Coulomb interaction weakens multifractality, e.g. for
$\epsilon=1/9$ we find $\Delta_2^{({\rm n})}=-0.48$ versus $\Delta_2=-0.047$ (see Table 2
). A qualitative reason for this is ``localizing" effect of the Coulomb interaction in the absence of time reversal and spin rotational symmetries. 
In the presence of interaction the transition occurs at smaller values of disorder, $t_*\ll t_*^{({\rm n})}$, which results in weakening of multifractality.

\textsf{Mesoscopic fluctuations of $\rho(E)$ above $T_c$.}\ --- Equation \eqref{eqmqRG-sym1} demonstrates that the anomalous dimensions of the moments of local density of states are affected by the interaction in the Cooper channel. Since the Cooper channel interaction diverges as the system approaches the temperature of superconducting transition $T_c$, one can expect enhancement of mesoscopic fluctuations of $\rho(E)$ in this case. To illustrate this effect we consider the case of weak short-ranged interaction in the particle-hole and particle-particle channels. Also, we assume that in the absence of interaction the system undergoes Anderson transition. Provided attraction in the Cooper channel dominates repulsion in the particle-hole channel, the system adjusts itself to the line $\gamma_s = - \gamma_t = - \gamma_c \equiv -\gamma$ under the renormalization group flow \cite{BurmistrovGornyiMirlin2012}. The corresponding renormalization group equation for $|\gamma|\ll 1$ and $|t-t_*^{({\rm n})}|\ll 1$ can be written as follows:
\begin{equation}
\frac{d t}{dy} = \nu^{-1}_{\rm n}\bigl (t-t_*^{({\rm n})}\bigr ) + \eta \gamma . \label{eq:AT:e4.2a}
\end{equation}
The second term in the right hand side of Eq. \eqref{eq:AT:e4.2a} describes the interaction correction to the conductivity.
In turn, the renormalization of $\gamma$ is described by the following equation:
\begin{equation}
\label{eq:AT:e4.1}
\frac{d\gamma}{ dy} = -\Delta_2^{\rm n} \gamma - a \gamma^2 .
\end{equation}
Here the constant $a$ is a universal number which is determined by the properties of composite operators at the noninteracting fixed point. We assume that the superconducting instability occurs by means of a standard BCS scenario, i.e. $a>0$. 
Eqs. \eqref{eq:AT:e4.2a} and \eqref{eq:AT:e4.1} allow one to estimate the transition temperature for case of initially weak interaction, $|\gamma|\ll 1$. In this case, we neglect the second term in the right hand side of Eq. \eqref{eq:AT:e4.1} and find
$\gamma(L)=\gamma e^{-\Delta_2^{\rm n} y}$. At finite $T$, the renormalization group flow is stopped at the length scale $L_T$. [We note that in this case $z=d$.] Estimating the transition temperature from the condition $|\gamma(L_{T})| \sim 1$, we find \cite{FeigelmanYuzbashyan2007,BurmistrovGornyiMirlin2012}:
\begin{equation}
T_c^* \sim \tau^{-1}\ |\gamma_0|^{d/|\Delta_2^{\rm n}|} ,
\label{eq:tcc}
\end{equation}
where $\tau$ denotes the mean free time.
%
%
\begin{table*}[t]
\begin{center}
\caption{Table 2. Anderson transitions in $d=2+\epsilon$ with and without Coulomb interaction. The 
value of $z$ in the interacting case has been obtained in Refs. \cite{baranov99,baranov02}. \label{Tabb2}}
\begin{tabular}{c|c}
interacting & non-interacting \\
\hline 
$t_*=\epsilon(1-A\epsilon)+O(\epsilon^3)$ & $t_*^{({\rm n})}=(2\epsilon)^{1/2}(1-3\epsilon/4)+O(\epsilon^{5/2})$ \\
$\nu=1/\epsilon-A+O(\epsilon)$ & $\nu_{\rm n}=1/2\epsilon-3/4+O(\epsilon)$ \\
$z=2+\epsilon/2+(2A-\pi^2/6-3)\epsilon^2/4+O(\epsilon^3)$ & $z_{\rm n}=2+\epsilon$ \\
$\theta = 1+O(\epsilon)$ & $\theta_{\rm n} =0$ \\
$\Delta_q = \frac{q(1-q)\epsilon}{4}\Bigl [1 + \left (1-A-\frac{\pi^2}{12}\right ) \epsilon \Bigr ]+O(\epsilon^3)$ & $\Delta_q^{\rm n} = q(1-q) \left (\frac{\epsilon}{2}\right )^{1/2} - \frac{3 \zeta(3)}{32} q^2(q-1)^2 \epsilon^2  +  O(\epsilon^{5/2})$ \\
& $z_\phi^{\rm n} =2+\sqrt{2\epsilon} +O(\epsilon)$ \\
\end{tabular}
\end{center}
\end{table*}

For $t<t_*^{({\rm n})}$ the renormalization group flows toward the metallic phase. After the length scale $\xi_{\rm n} = l |t/t_*^{({\rm n})}-1|^{-\nu_{\rm n}}$, the disorder-induced renormaization of $\gamma$ is stopped and the standard disorder-free BCS mechanism with 
attraction $\gamma(\xi_{\rm n}) = \gamma_{0} (\xi_{\rm n}/l)^{|\Delta_2^{\rm n}|}$ yields the superconducting instability at temperature \cite{BurmistrovGornyiMirlin2012}:
\begin{equation}
T_c(\xi_{\rm n}) = \delta_\xi e^{-1/|\gamma(\xi_{\rm n})|}
= \delta_\xi \exp \left [ - \frac{a}{|\Delta_2^{\rm n}|} \left (\frac{T_c^*}{\delta_\xi}\right )^{\Delta_2^{\rm n}/d} \right ] .
\label{eq:AT:e4.7}
\end{equation}
Here $\delta_\xi = \tau^{-1}(\xi_{\rm n}/l)^{-d}$ denotes the typical level spacing in a volume of size $\xi_{\rm n}$. It is analogous to the scale $\Delta_\xi$ since for non-interacting critical point $z_{\rm n} = d$.
Eq. \eqref{eq:AT:e4.7} interpolates between $T_c^*$ at $\delta_\xi
\sim T_c^*$  and $T_c^{\rm BCS}= \tau^{-1}\exp(-1/|\gamma|)$ at $\delta_\xi \sim 1/\tau$.

\begin{figure}[t]
\centerline{\includegraphics[width=0.9\columnwidth]{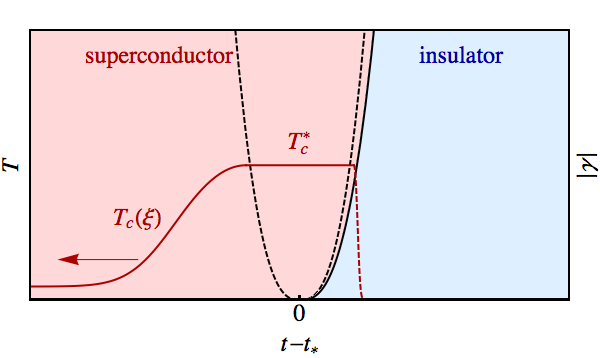}}
\caption{Fig. \protect\ref{Figure4}. Sketch of the phase diagram in disorder ($t$) and interaction ($|\gamma|$) plane near the superconductor-insulator transition. The solid black curve denotes the transition. The dashed black curve corresponds to the condition $T_c^*\sim \delta_\xi$ and indicates the critical region. The red solid curve illustrates the dependence of the superconducting transition temperature on the distance from the critical point. 
}
\label{Figure4}
\end{figure}

For $t>t_*^{({\rm n})}$ the non-interacting system is in the insulating phase. In the presence of interaction there are two possibilities. For $\delta_\xi<T_c^*$ $|\gamma|$ approaches unity in the critical region. Then the superconducting phase is expected to be established at  $T<T_c^*$. For $\delta_\xi>T_c^*$ $t$ becomes unity while $|\gamma|\ll 1$. Thus in this case one can expect localization. Therefore we can estimate position of the quantum phase transition between superconducting and insulating phases as follows $\delta_\xi \sim T_c^*$. The latter is equivalent to the following relation: $|\gamma| \sim \bigr (t_*^{({\rm n})}-t\bigr )^{\nu_{\rm n}|\Delta_2^{\rm n}|}$ (see Fig. \ref{Figure4}). We note that in Ref.~\cite{FeigelmanCuevas2010} superconducting state with $T_c \ll T_c^*$ was found to survive in the localized regime, $\delta_\xi > T_c^*$,  due to Mott-type rare configurations. 

Within plain perturbation theory the average density of states near $T_c$ is strongly affected by Cooper channel attraction \cite{Abrahams1970,Maki1970,Castro1990}. These classical results can be extended to incorporate the renormalization group flow near the non-interacting critical point \cite{Burmistrov2016}.

For $|t-t_*^{({\rm n})}|\ll 1$ the anomalous dimension governing scaling behaviour of the $q$-th moment of the local density of states DOS can be written in the form of series expansion in $\gamma$:
\begin{equation}
\zeta_q = \Delta_q^{\rm n} - b_q \gamma .
\label{eq:mq:rg:fp}
\end{equation}
Here $b_q$ are some universal coefficients characterizing critical behaviour at the non-interacting fixed point. 
The following comments are in order here: (i) the expansion of $\zeta_q$ in interaction parameter is consistent with the expansion in powers of $t$ (see Table 1
); (ii) in the case of short-ranged interaction one needs to distinguish between the scales $L_E \sim |E|^{-1/d}$ and $L_T \sim T^{-1/d}$ on the one hand, and the dephasing length $L_\phi \sim \tau_{\phi}^{1/d}$, on the other hand. The dephasing time $\tau_\phi$ is expected to have a power-law dependence on energy and temperature, $\tau_\phi \sim (\max\{|E|,T\})^{-p}$. Interaction correction in Eq. \eqref{eq:mq:rg:fp} is stopped at the scale $\min\{L_E, L_T\}$ whereas the non-interacting renormalization is extended up to the dephasing length $L_\phi$. Typically, the following condition holds, $L_\phi \gg L_E, L_T$.

On the metallic side of the transition (including the critical point), $t\leqslant t_*^{({\rm n})}$, Eqs. \eqref{eq:AT:e4.1} and \eqref{eq:mq:rg:fp} imply
the following result for the moments of the local density of states:      
\begin{equation}
\frac{\langle \rho^q(E)\rangle}{\langle \rho(E)\rangle^q} = \left ( \frac{\mathcal{L}}{L_\phi}\right )^{\Delta_q^{\rm n}}\left ( \frac{\gamma(\mathcal{L})}{\gamma_0}\right )^{\Delta_q^{\rm n}/\Delta_2^{\rm n}}
\left ( \frac{\Delta_2^{\rm n}+a \gamma_0}{\Delta_2^{\rm n}+a \gamma(\mathcal{L})}\right )^{x_q^{\rm n}} ,
\end{equation}
where $x_q=b_q/a+ \Delta^{\rm n}_q/\Delta^{\rm n}_2$ and $\mathcal{L}=\min\{\xi_{\rm n},L_E, L_T\}$.
For $|E|, T \gg T_c(\xi)$ interaction at the scale $\mathcal{L}$ is small, $|\gamma(\mathcal{L})|\ll 1$. Then from Eq. \eqref{eq:AT:e4.1} we find  $\gamma(\mathcal{L}) \sim \mathcal{L}^{-\Delta_2^{\rm n}}$. Therefore, the moments of the local density of states are scaled in the same way as  in the absence of interaction: ${\langle \rho^q(E)\rangle}/{\langle \rho(E)\rangle^q} \sim L_\phi^{-\Delta_q^{\rm n}}$.

At criticality, $\delta_\xi\ll T_c^*$, and in the vicinity of transition temperature, $T-T_c^*\ll T_c^*$, the moments of the local density of states are enhanced significantly for energies $|E| \ll T_c^*$, due to divergence of $\gamma(\mathcal{L})$ at $\mathcal{L} = L_{T_c^*}$:
\begin{equation}
\frac{\langle \rho^q(E)\rangle}{\langle \rho(E)\rangle^q} = \left ( \frac{L_{T_c^*}}{L_\phi}\right )^{\Delta_q^{\rm n}}\left ( \frac{\gamma(L_T)}{\gamma_0}\right )^{-b_q/a}
,
\end{equation}
where $L_\phi$ is evaluated at $T=T_c^*$.  On the insulating side at $\delta_\xi \gtrsim T_c^*$ we find standard insulating behaviour for the non-interacting critical point:
\begin{equation}
\frac{\langle \rho^q(E)\rangle}{\langle \rho(E)\rangle^q}  =
\bigl ( \min\{L_\phi,\xi_{\rm n}\}\bigr )^{-\Delta_q^{\rm n}} \left (\frac{\max\{\xi_{\rm n},L_\phi\}}{\xi_{\rm n}}\right )^{d(q-1)} .
\label{111}
\end{equation}

\textsf{Conclusions.}\ --- In conclusion, we reviewed our recent results for mesoscopic fluctuations of the local density of states in the presence of electron-electron interaction. Specifically, we focused on two cases: (i) a vicinity of Anderson-Mott transition and (ii) vicinity of the non-interacting critical point in the presence of a weak electron-electron attraction. 

For Mott-Anderson transition we found that the strong mesoscopic fluctuations (multifractality) of the local density of states survive in the presence of electron-electron interaction. Within two-loop expansion in disorder we check that the multifractal spectrum in the presence of interaction is different from the multifractal spectrum known in the absence of interaction. In addition, we demonstrated that in some cases on the insulating side of Anderson-Mott transition the mobility edge for single particle excitations can exist. The mobility edge has a nontrivial scaling with the distance from the interacting critical point.
We note that many-body delocalization driven by long-range (Coulomb) interaction
may affect the localization transition at $E_c$  \cite{gutman2016}. The detailed discussion of this issue is beyond the scope of the present paper.

For the case of vicinity of the non-interacting critical point in the presence of weak electron-electron attraction we found enhancement of mesoscopic fluctuations at temperatures close to the superconducting transition temperature, $T-T_c^*\ll T_c^*$. At high temperatures the mesoscopic fluctuations of $\rho(E)$ are the same as in the non-interacting case.

The predicted strong mesoscopic fluctuations of local density of states imply strong point-to-point fluctuations of tunneling spectra which can be measured in scanning tunneling microscopy experiments. We note that our theoretical results are consistent with available data on scanning tunneling microscopy in disordered interacting systems, in particular, for
a strongly disordered 3D system \cite{morgenstern02}, for various 2D semiconductor systems and graphene \cite{morgenstern-2D-1,morgenstern-2D-2,morgenstern-2D-3,morgenstern-2D-4}, for a magnetic semiconductor Ga$_{1-x}$Mn$_x$As near metal-insulator transition \cite{Richardella}, for metallic and insulating phases near superconductor-insulator transition in TiN, InO, and NbN films \cite{Sacepe08,Sacepe10,Sacepe11,Sherman14,Mondal11,Noat13}.

Finally, we mention that the moments of the local density of states in the presence of interaction are represented as pure-scaling local operators of Finkel'stein nonlinear sigma model. Recently, in the presence of electron-electron interaction the wide set of pure-scaling local operators has been constructed by one of us with Repin \cite{Repin2016}. This set of pure-scaling local operators is a generalization of operators constructed for non-interacting case \cite{Wegner1986,Wegner1987a,Wegner1987b}.

The authors are grateful to I. Gruzberg, D. Gutman, M. Feigel'man, V. Kravtsov, D. Lyubshin, A. Ludwig, M. M\"uller, P. Ostrovsky, D. Polyakov, I. Protopopov, B. Shklovskii, M. Skvortsov, K. Tikhonov, and A. Yazdani
for stimulating discussions. I.V.G. and A.D.M. acknowledge support by the Deutsche Forschungsgemeinschaft (grants MI 658/9-1 and GO 1405/3-1).

\end{document}